\documentstyle[prd,aps]{revtex}

\newcommand{\D}{{\cal D}}

\newcommand{\be}{\begin{equation}}
\newcommand{\ee}{\end{equation}}
\newcommand{\ba}{\begin{eqnarray}}
\newcommand{\ea}{\end{eqnarray}}
\newcommand{\bi}{\begin{itemize}}
\newcommand{\ei}{\end{itemize}}

\newcommand{\RR}{{\rm I\kern -.2em  R}}

\begin{document}

\title{WHITENING OF THE QUARK-GLUON PLASMA}

\author{Cristina Manuel\footnote{Electronic address:
{\tt cristina.manuel@ific.uv.es}}}

\address{\it Instituto de F\'{\i}sica Corpuscular \\
C.S.I.C.-Universitat de Val\`encia\\
Edificio de Institutos de Paterna, Apt 2085 \\
46071 Val\`encia, Spain}

\author{Stanis\l aw Mr\' owczy\' nski\footnote{Electronic address:
{\tt mrow@fuw.edu.pl}}}

\address{\it So\l tan Institute for Nuclear Studies \\
ul. Ho\.za 69, PL - 00-681 Warsaw, Poland \\
and Institute of Physics, \'Swi\c etokrzyska Academy \\
ul. \'Swi\c etokrzyska 15, PL - 25-406 Kielce, Poland}

\date{31-st August 2004}

\maketitle

\begin{abstract}

Parton-parton collisions do not neutralize local color charges in
the quark-gluon plasma as they only redistribute the charges among
momentum modes. We discuss color diffusion and color conductivity
as the processes responsible for the neutralization of the plasma.
For this purpose, we first compute the conductivity and diffusion
coefficients in the plasma that is significantly colorful. Then,
the time evolution of the color density due to the conductivity and 
diffusion is studied. The conductivity is shown to be much more 
efficient than the diffusion in neutralizing the plasma at the scale 
longer than the screening length. Estimates of the characteristic 
time scales, which are based on close to global equilibrium computations, 
suggest that first the plasma becomes white and then the momentum 
degrees of freedom thermalize.

\end{abstract}

\pacs{PACS: 12.38.Mh, 05.20.Dd, 11.10.Wx}



\section{Introduction}


Production of the quark-gluon plasma is expected at the early stage
of high-energy nucleus-nucleus collision when the energy density is
sufficiently high. The experimental data on the so-called elliptic flow
\cite{Ackermann:2000tr}, which have been obtained at the Relativistic
Heavy-Ion Collider (RHIC) in Brookhaven National Laboratory, suggest
a surprisingly short, below 1 ${\rm fm}/c$ \cite{Heinz:2001xi},
equilibration time of the system. Understanding of the thermalization
process is thus a key issue of the quark-gluon plasma physics.

In our previous study \cite{Manuel:2003zr}, we have analyzed the local
equilibrium of the plasma, which is defined as a state of maximal local
entropy. Using the kinetic equations with the collision terms of the
Waldmann-Snider form, we have proved that such a state is generically
{\em colorful}, {\it i.e.} the color four-current is non-vanishing. Thus,
the collisions, which are responsible for equilibration of the parton
momenta, do {\em not} neutralize the local color density. Since the color
current is (covariantly) conserved in every collision process, the
inter-parton collisions redistribute the color charges among various
momentum modes but they do not change a local macroscopic color charge.
Consequently, if the color charges are not homogeneously distributed
in the process of the plasma production due to, say, statistical
fluctuations, the inter-parton collisions will not neutralize the system.
On the other hand, the global equilibrium of the quark-gluon plasma is
locally colorless because of the maximum entropy principle. We assume
here that the system does not carry a global color charge and that
it does not experience an external chromodynamic field. Once the inter-parton
collisions are not responsible for the neutralization of the local charges,
one has to invoke other collective mechanism to whiten the plasma.
This is the subject of this article.

Local charges are neutralized due to the currents that flow in the system.
We consider the diffusive currents generated by the charge density
gradient (Fick's law) and the ohmic currents caused by the chromoelectric
field (Ohm's law) which is, in turn, induced by the charge density.
The color conductivity of the quark-gluon plasma has been studied
for long time
\cite{Heinz:qe,Czyz:1986mr,Dyrek:1986vv,Mrowczynski:xu,Mrowczynski:1989bv,Selikhov:1993ns,Selikhov:xn,Heiselberg:px},
but only recently the problem has been well understood
\cite{Bodeker:1998hm,Bodeker:1999ey,Litim:1999ns,Litim:1999id,Arnold:1998cy,Blaizot:1999xk,Arnold:1999uy,MartinezResco:2000pz}.
As far as we know, the color diffusion was only briefly discussed in
\cite{Heiselberg:px}. In all these papers, the plasma near the colorless
global equilibrium was studied. We are, however, interested in the plasma
that is locally colorful. Thus, in Sec.~\ref{D-sigma-constants} we derive
the diffusion and conductivity coefficients in such a plasma, and then, in
Sec.~\ref{t-evolution} the temporal evolution of the color charge density
is considered. The ohmic currents are shown to be much more efficient than
the diffusive ones in neutralizing the local charges.

If not stated otherwise, we follow here the same conventions and notations
as in \cite{Manuel:2003zr}.


\section{Color Diffusion and conductivity coefficients}
\label{D-sigma-constants}


In this section we derive, using  transport theory, the diffusive and
ohmic currents  in a plasma that is locally colorful. The transport
equations of quarks, antiquarks and gluons, which form the basis of
our analysis, read
\begin{mathletters}
\ba
\label{q-transport-eq1}
\big(D^0 + {\bf v} \cdot {\bf D} \big) Q({\bf p},x)
- {g \over 2}
\{{\bf E} + {\bf v} \times {\bf B}, \nabla_p Q({\bf p},x) \}
&=&  C[Q,\bar Q,G] \;, \\ [2mm]
\label{aq-transport-eq1}
\big(D^0 + {\bf v} \cdot {\bf D} \big) \bar Q({\bf p},x)
+ {g \over 2}
\{{\bf E} + {\bf v} \times {\bf B}, \nabla_p \bar Q({\bf p},x) \}
&=&  \bar C[Q,\bar Q,G] \;, \\ [2mm]
\label{g-transport-eq1}
\big(D^0 + {\bf v} \cdot {\bf D} \big) G({\bf p},x)
- {g \over 2}
\{{\bf E} + {\bf v} \times {\bf B}, \nabla_p G({\bf p},x) \}
&=&  C_g[Q,\bar Q,G] \;.
\ea
\end{mathletters}
The (anti-)quark on-mass-shell distribution functions $Q({\bf p},x)$
and $\bar Q({\bf p},x)$, which are $N_c\times N_c$ hermitean
matrices, belong to the fundamental representation of the SU($N_c$)
group, while the gluon distribution function $G({\bf p},x)$,
which is a $(N_c^2 -1) \times (N_c^2 -1)$ matrix, belongs to
the adjoint representation. The covariant derivative
$D^\mu \equiv \partial^\mu - ig [A^\mu(x),\cdots \;]$, the
chromoelectric and chromomagnetic fields, ${\bf E}$ and ${\bf B}$,
which enter the transport equations also belong to either the
fundamental or adjoint representation, correspondingly.
To simplify the notation, and differently than in \cite{Manuel:2003zr},
we use the same symbols $D^0$, ${\bf D}$, ${\bf E}$ and
${\bf B}$ to denote a given quantity in the fundamental or adjoint
representation. $x \equiv (t,{\bf x})$ denotes the four-position
while ${\bf p}$ is the three-momentum. Because the partons are
assumed to be massless, the velocity ${\bf v}$ equals
${\bf p}/|{\bf p}|$. The collision terms $C$, $\bar C$ and
$C_g$ will be discussed later on.

We are interested in a state  close to the colorful local
equilibrium. When the effects of quantum statistics are neglected,
the (on-mass-shell) local equilibrium distribution functions read
\cite{Manuel:2003zr}
\begin{mathletters}
\label{loc-eq}
\ba \label{loc-eq-Q}
Q^{\rm eq}({\bf p},x) &=& {\rm exp}\Big[- \beta (x)
\big(u_\nu (x) p^\nu - \mu_b(x) - \widetilde \mu (x) \big)\Big] \;,
\\ [2mm] \label{loc-eq-barQ}
\bar Q^{\rm eq}({\bf p},x) &=& {\rm exp}\Big[-\beta (x)
\big(u_\nu (x) p^\nu + \mu_b(x) + \widetilde \mu (x) \big)\Big] \;,
\\ [2mm] \label{loc-eq-G}
G^{\rm eq}({\bf p},x) &=& {\rm exp}\Big[-\beta (x)
\big(u_\nu (x) p^\nu - \widetilde \mu_g (x) \big)\Big] \;,
\ea
\end{mathletters}
where $p^\mu = (E_p,{\bf p})$, and $E_p =|{\bf p}|$ for massless quarks
and antiquarks, and for gluons; $\beta = 1/T$, $u^\mu$, $\mu_b$ denote,
respectively, the inverse temperature, hydrodynamic velocity and baryon
chemical potential; the colored chemical potentials of quarks
$(\widetilde \mu)$ and of gluons $(\widetilde \mu_g)$ obey
the relationship
\be
\label{relat-qgmu}
\widetilde \mu _g (x)= 2 T^a {\rm Tr}[\tau^a \widetilde \mu(x) ]
= \mu_a (x)  T^a \;,
\ee
where $\tau^a$, $T^a$ with $a = 1, ... \, ,N_c^2-1$ are the SU($N_c$)
group generators in the fundamental and adjoint representations,
normalized as ${\rm Tr}[\tau^a \tau^b] = \frac12 \delta^{ab}$ and
${\rm Tr}[T^a T^b] = N_c \delta^{ab}$.

In the local equilibrium state there is a non-vanishing color charge
density, which is defined as
\be
\rho(x) = - {g \over 2} \int {d^3 p \over (2\pi)^3} \,
\Big( q({\bf p},x) - \bar q({\bf p},x)
+ 2 \tau^a {\rm Tr}\big[T^a G({\bf p},x) \big] \Big) \;,
\ee
with
$$
q({\bf p},x) \equiv Q({\bf p},x)
- {1 \over N_c} {\rm Tr}\big[Q({\bf p},x)\big] \;, \;\;\;\;\;\;
\bar q({\bf p},x) \equiv Q({\bf p},x)
- {1 \over N_c} {\rm Tr}\big[\bar Q({\bf p},x)\big] \;.
$$
In the local rest frame, where $u^\mu = (1,0,0,0)$, the color current
defined as
\be
{\bf j}(x) = - {g \over 2} \int {d^3 p \over (2\pi)^3} \, {\bf v}
\Big( q({\bf p},x) - \bar q({\bf p},x)
+ 2 \tau^a {\rm Tr}\big[T^a G({\bf p},x) \big] \Big) \;,
\ee
vanishes because the distribution functions are locally isotropic.

We  now study the system for long time scales.
We consider small deviations from local equilibrium, and
we write down the quark distribution function as
$Q({\bf p},x) =  Q^{\rm eq}({\bf p},x) + \delta Q({\bf p},x)$.
Assuming that
\be
\label{q-conditions}
|Q^{\rm eq}| \gg |\delta Q| \;, \;\;\;\;\;\;\;\;
|D^0 Q^{\rm eq}| \gg |D^0 \delta Q| \;, \;\;\;\;\;\;\;\;
|{\bf D} Q^{\rm eq}| \gg | {\bf D} \delta Q| \;, \;\;\;\;\;\;\;\;
|\nabla_p Q^{\rm eq}| \gg | \nabla_p \delta Q| \;,
\ee
and taking into account the local isotropy of the equilibrium state,
the transport equation (\ref{q-transport-eq1}) can be approximated as
\be
\label{q-transport-eq2}
\left(D^0 +{\bf v} \cdot {\bf D}\right) Q^{\rm eq} +
{g \over 2} \{{\bf E}, \nabla_p Q^{\rm eq}\}
= L[\delta Q] \;,
\ee
where $L[\delta Q]$ is the collision term linearized around the
local equilibrium distribution function. Analogous equations hold
for antiquarks and gluons. Let us recall here that the collision
terms evaluated for the local equilibrium distribution functions
(\ref{loc-eq}) vanish \cite{Manuel:2003zr}.

To get the transport coefficients of color diffusion and
conductivity we assume that the linearized collision terms $L$,
$\bar L$ and $L_g$ satisfy the relationship 
\be 
\label{trick} 
\int {d^3 p \over (2\pi)^3} \, {\bf v} \, L[\delta Q] = - \gamma \int
{d^3 p \over (2\pi)^3} \, {\bf v}\, \delta Q \;, 
\ee 
where $1/\gamma$ is the characteristic relaxation time which, for
simplicity, is assumed to be the same for quarks, antiquarks and gluons. 
Our analysis based on the linearized transport equation (\ref{q-transport-eq2}) 
with Ansatz (\ref{trick}) is valid for $t \gtrsim 1/\gamma$. The relation 
(\ref{trick}) is obeyed \cite{Arnold:1998cy} by the Waldmann-Snider collision 
term linearized around global (colorless) equilibrium. Such a linear collision 
term turns out to be non-local in velocities, allowing for the covariant color
current conservation \cite{Bodeker:1998hm}. A similar linearization around 
the local colorful equilibrium is, in principle, feasible, but it appears 
rather difficult as it requires knowledge of the scattering matrix elements 
computed in the colorful background.

The relationship (\ref{trick}) is trivially satisfied
by the collision term in the relaxation time approximation
\be
\label{RTA}
C^{\rm RTA}[Q,\bar Q, G] = {1 \over \tau}
\big(Q^{\rm eq} ({\bf p},x) - Q({\bf p},x) \big)
= - {1 \over \tau} \, \delta Q ({\bf p},x) = L^{\rm RTA}[\delta Q]\;,
\ee
with $\tau = 1/\gamma$. Unfortunately, this approximation is known
to contradict the covariant current conservation as $C$, $\bar C$
and $C_g$ in the form (\ref{RTA}) violate the collisional invariant
\be \label{col-cons-inv}
\int {d^3 p \over (2\pi)^3}  \;
\big( C - \bar C + 2 \tau^a {\rm Tr}[ T^a C_g] \big) = 0 \;.
\ee
However, the collision term (\ref{RTA}) can be improved to a form that
complies with the condition (\ref{col-cons-inv}). In analogy to the
so-called BGK collision term \cite{BGK}, we have found the expression
\ba
\label{improved-RTA}
C^{\rm BGK}[Q,\bar Q, G] &=& {1 \over \tau} \Big(Q({\bf p},x) -
N(x) \, \big(N^{\rm eq}(x)\big)^{-1} Q^{\rm eq}({\bf p},x) \Big)
\\ [2mm] \nonumber
&\approx & - {1 \over \tau} \Big(\delta Q ({\bf p},x)
- \delta N(x) \, \big(N^{\rm eq}(x)\big)^{-1}
Q^{\rm eq}({\bf p},x) \Big) = L^{\rm BGK} [\delta Q]\;,
\ea
with
\be
N(x) \equiv \int {d^3 p \over (2\pi)^3} \, Q({\bf p},x) \;,
\ee
and $(N^{\rm eq})^{-1}$ being the inverse matrix of $N^{\rm eq}$.
One easily shows that the collision terms of the form
(\ref{improved-RTA}) satisfy the constraint (\ref{col-cons-inv}), as
$$
\int {d^3 p \over (2\pi)^3}  \, C^{\rm BGK}
=\int {d^3 p \over (2\pi)^3}  \, \bar C^{\rm BGK} =
\int {d^3 p \over (2\pi)^3}  \, C_g^{\rm BGK} = 0   \;.
$$
The collision term (\ref{improved-RTA}) also obeys the relation
(\ref{trick}). It still violates the energy-momentum conservation law,
but it can be further
improved \cite{BGK}.

Using the relation (\ref{trick}), the color current  generated
by deviations from equilibrium is
\ba
\label{q-current2}
{\bf j}(x) &=& - {g \over 2} \int {d^3 p \over (2\pi)^3} \, {\bf v}
\Big( \delta q({\bf p},x) -  \delta \bar q({\bf p},x)
+ 2 \tau^a {\rm Tr}\big[T^a\delta G({\bf p},x)) \big] \Big)
\\ [2mm] \nonumber
&=& {g \over 2\gamma} D^0 \int {d^3 p \over (2\pi)^3} \, {\bf v}
\Big(q^{\rm eq} - \bar q^{\rm eq}
+  2 \tau^a {\rm Tr}\big[T^a G^{\rm eq} \big] \Big)
\\  [2mm] \nonumber
&+& {g \over 2\gamma} \int {d^3 p \over (2\pi)^3} \, {\bf v}
( {\bf v} \cdot {\bf D})
\Big(q^{\rm eq} - \bar q^{\rm eq}
+  2 \tau^a {\rm Tr}\big[T^a G^{\rm eq} \big] \Big)
\\  [2mm] \nonumber
&-& {g^2 \over 4\gamma} \int {d^3 p \over (2\pi)^3} \, {\bf v}
\Big( \{{\bf E}, \nabla_p (Q^{\rm eq} + \bar Q^{\rm eq})\}
- {2 \over N_c} {\rm Tr}\big[{\bf E} \cdot
\nabla_p (Q^{\rm eq} + \bar Q^{\rm eq}) \big] \Big)
\\  [2mm] \nonumber
&-& {g^2 \over 2\gamma}  \tau^a \int {d^3 p \over (2\pi)^3} \, {\bf v} \,
{\rm Tr}[T^a \{ {\bf E},\nabla_p G^{\rm eq} \} ] \;.
\ea
One observes that the term with $D^0$ drops out as the color
current vanishes in local equilibrium. We also take into account
that the local equilibrium is isotropic and we perform partial
integrations in the terms with the chromoelectric field. Putting
additionally ${\bf v}^2=1$, the current gets the form
\ba
{\bf j} &=& {g \over 6\gamma} \, {\bf D}
\int {d^3 p \over (2\pi)^3} \Big(q^{\rm eq} - \bar q^{\rm eq}
+ 2 \tau^a {\rm Tr}\big[T^a G^{\rm eq} \big] \Big)
\\  [2mm] \nonumber
&+& {g^2 \over 6\gamma}
\int {d^3 p \over (2\pi)^3} \: {1 \over E_p}
\Big( \{{\bf E}, (Q^{\rm eq} + \bar Q^{\rm eq}) \}
- {2 \over N_c} {\rm Tr}[{\bf E} \,  (Q^{\rm eq} + \bar Q^{\rm eq})] \Big)
\\  [2mm] \nonumber
&+& {g^2 \over 3\gamma} \,  \tau^a \int {d^3 p \over (2\pi)^3}
\: {1 \over E_p} {\rm Tr}[T^a \{{\bf E}, G^{\rm eq} \}] \;,
\ea
which can be written as
\be
\label{current-RTA}
{\bf j} = - {\cal D} {\bf D} \rho + {1 \over 2}
\big( \{\sigma_q, {\bf E}\}
- {2 \over N_c} {\rm Tr}[\sigma_q \, {\bf E}] \big)
+ \tau^a  {\rm Tr}[T^a \{ \sigma_g , {\bf E} \}]  \;,
\ee
where the diffusion constant ${\cal D}$ and conductivity coefficients
$\sigma_q$, $\sigma_g$ are
\be
\label{D-sigma-RTA}
{\cal D} = {1 \over 3\gamma}
\;, \;\;\;\;\;\;\;
\sigma_q = {g^2 \over 3\gamma}
\int {d^3 p \over (2\pi)^3} \, {1 \over E_p}
\, \Big( Q^{\rm eq} + \bar Q^{\rm eq} \Big)
\;, \;\;\;\;\;\;\;
\sigma_g = {g^2 \over 3\gamma}
\int {d^3 p \over (2\pi)^3} \, {1 \over E_p} \, G^{\rm eq} \;.
\ee
As seen, the current (\ref{current-RTA}) is traceless as it
should. It gets a much simpler form in the adjoint representation.
Namely, for ${\bf j}^a = 2 {\rm Tr}[\tau^a {\bf j}]$, we get
\be \label{adj-current}
{\bf j}^a(x) = - {\cal D}\, {\bf D}^{ab} \rho^b(x)
+ \sigma^{ab} {\bf E}^b(x)   \;,
\ee
where ${\bf D}^{ab} = \delta^{ab} {\bf \nabla} + g f^{acb} {\bf A}^c$,
and the color conductivity tensor reads
\be
\label{cctensor}
\sigma^{ab} = {g^2 \over 3 \gamma}
\int {d^3 p \over (2\pi)^3} \, {1 \over E_p} \,
\Big( {\rm Tr} [\{\tau^a,\tau^b\} ( Q^{\rm eq} + \bar Q^{\rm eq})]
+ {\rm Tr} [\{T^a,T^b\} G^{\rm eq}] \Big) \;.
\ee
The diffusion constant is, as previously, $1/3\gamma$. When the
equilibrium is colorless, the conductivity $\sigma$ is proportional
to the unit matrix in color space, but for a colorful configuration
it is not.

It is interesting to note that the chromoelectric field contributes
to the induced baryon current. Repeating the analysis fully analogous
to that of the color current, one finds that the baryon current
defined as
\be
\label{b-current-def}
{\bf b}(x) =  \frac 13 \int {d^3 p \over (2\pi)^3} \, {\bf v}\,
{\rm Tr}[ Q({\bf p},x) - \bar  Q({\bf p},x)]  \;.
\ee
equals
\be
\label{b-current-RTA}
{\bf b}(x) = - {\cal D} \nabla b (x)
- {2 \over 3g} {\rm Tr}[\sigma_q \, {\bf E}(x)] \;,
\ee
where $b$ is the baryon density while ${\cal D}$ and $\sigma_q$
are given by Eqs.~(\ref{D-sigma-RTA}). When the equilibrium
is colorless, the conductivity $\sigma$ is proportional to the
unit matrix in color space, and the effect of the chromoelectric
field on the baryon current disappears.


\section{Temporal evolution of the color density}
\label{t-evolution}


Our aim here is to discuss how a locally colorful quark-gluon plasma
becomes white. As an introduction to our chromodynamic considerations,
we first discuss the temporal evolution of the electric charge density in
the electromagnetic plasma.


\subsection{Diffusion vs. conductivity - electrodynamic case}


The electric current (${\bf j}$) generated by both the gradient of
charge density ($\rho$) and the electric field (${\bf E}$) is
\be 
\label{diff-cond-curr}
{\bf j}(x) = - \D \, \nabla \rho (x) + \sigma \, {\bf E}(x) \;.
\ee
$\D$ and $\sigma$ are assumed here to be the transport coefficients 
derived in a semi-static limit as in Sec.~\ref{D-sigma-constants}. 
Therefore, Eq.~(\ref{diff-cond-curr}) holds for slowly varying
$\rho (x)$ and ${\bf E}(x)$. 

Taking into account the Gauss law $\nabla {\bf E}(x) = \rho (x)$, 
the current conservation $\partial \rho/ \partial t + \nabla {\bf j} = 0$
combined with Eq.~(\ref{diff-cond-curr}) provides the equation
\be 
\label{diff-cond-eq}
\bigg[ {\partial \over \partial t} -  \D \,\nabla^2 + \sigma \bigg]
\rho (x) = 0 \;,
\ee
which, supplemented by the initial condition
$\rho(0,{\bf x}) = \rho_0({\bf x})$, is solved by
\be
\rho (x) = e^{-\sigma t} \, n(x)
\ee
with $n$ satisfying the diffusion equation
\be \label{diff-eq}
\bigg[ {\partial \over \partial t} - \D \,\nabla^2 \bigg]
n(x) = 0 \;,
\ee
and the initial condition $n(0,{\bf x}) = \rho_0({\bf x})$.
Eq.~(\ref{diff-cond-eq}) can be easily solved by means of the 
Fourier transformation as
\be
\label{solution}
\rho (x) = \int \frac{d^3 k}{(2\pi )^3} \, 
e^{-(\sigma + \D{\bf k}^2)t + i{\bf kx}}
\rho_0({\bf k}) \;,
\ee
where $\rho_0({\bf k})$ is the Fourier transform of the initial
charge density
\be
\label{Fourier}
\rho_0({\bf k}) = \int d^3x  \,
e^{- i{\bf kx}} \rho_0({\bf x}) \;.
\ee
It is assumed here that the integral (\ref{Fourier}) exsists which
requires vanishing of $\rho_0({\bf x})$ when $|{\bf x}| \rightarrow \infty$. 

We note that the solution (\ref{solution}) can be also written down as
\be
\label{solution2}
\rho (x) = {1 \over (4\pi \D t)^{3/2}}
\int d^3x' \: 
{\rm exp}\bigg(-\sigma t - {({\bf x}-{\bf x}')^2 \over 4 \D t} \bigg) 
\rho_0({\bf x}') \;,
\ee
where the Green's function  
\be
G({\bf x}, {\bf x}',t) = 
{1 \over (4\pi \D t)^{3/2}}
{\rm exp}\bigg(-\sigma t - {({\bf x}-{\bf x}')^2 \over 4 \D t} \bigg)
\;,
\ee
represents the charge density which obeys Eq.~(\ref{diff-cond-eq})
and equals $\delta^{(3)}({\bf x}- {\bf x}')$ at $t=0$. 

The solution (\ref{solution}) shows that the charge density modes  
of all ${\bf k}$ decay exponentially. The long-wavelength 
modes with ${\bf k}^2 < \sigma /\D$ are dominantly neutralized 
by the ohmic currents while those with ${\bf k}^2 > \sigma /\D$
are neutralized due to the diffusion. It should be remembered, 
however, we can trust the solution (\ref{solution}) or (\ref{solution2}) 
only for sufficiently long time intervals because Eq.~(\ref{diff-cond-curr}) 
holds for slowly varying $\rho (x)$ and ${\bf E}(x)$. We return 
to this point in the next section where it is discussed 
quantitatively in the context of quark-gluon plasma.


\subsection{Diffusion vs. conductivity - chromodynamic case}
\label{colordenev}


The color current generated by both the gradient of color density and
the chromoelectric field is given by Eq.~(\ref{adj-current}). The
covariant current conservation $D^0 \rho + {\bf D} {\bf j} = 0$
combined with the Gauss law ${\bf D} {\bf E}(x) = \rho (x)$,
provides the equation
\be
\label{col-diff-cond-eq}
\big[ D^0 - \D \, {\bf D}^2 + \sigma \big]
\rho (x) = 0 \;,
\ee
where the term $[{\bf D}, \sigma] \,{\bf E}$ has been neglected.
In the Appendix we show that
$[{\bf D}, \sigma] \,{\bf E} \ll \sigma {\bf D} {\bf E}$ in the small
coupling limit. All chromodynamic quantities discussed in this
section belong to the adjoint representation, and thus, the color
indices are suppressed.

Eq.~(\ref{col-diff-cond-eq}) can be treated as its Abelian
counterpart (\ref{col-diff-cond-eq}) if $[D^0,\sigma] = 0$.
In the Appendix this commutator is shown to be indeed small.
Thus, Eq.~(\ref{col-diff-cond-eq}) is solved by
\be
\label{qgp-solution}
\rho (x) = e^{-\sigma t} \, n(x)
\ee
with $n$ satisfying the diffusion equation
\be 
\label{col-diff-eq}
\big[ D^0 - \D \, {\bf D}^2 \big] n(x) = 0 \;.
\ee
$\rho (x)$ and $n(x)$ obey the initial condition
$\rho(0,{\bf x}) = n (0,{\bf x}) = \rho_0({\bf x})$. Thus, we expect 
that, as in the electromagnetic case, the charge density decays 
exponentially and the conductivity dominates over the diffusion 
for the modes with ${\bf k}^2 < \sigma /\D$.

For further discussion one needs an estimate of the relaxation
time $1/\gamma$ which controls both $\sigma$ and $\D$. However, the 
reliable estimate can be given only for the quark-gluon plasma close 
to global (colorless) equilibrium of very high temperatures where 
$1/g \gg 1$. Then, the color conductivity is of order 
\cite{Selikhov:1993ns,Arnold:1998cy}
\be
\label{sigma-est}
\sigma ~  \sim {T \over {\rm ln}(1/g) }  \;.
\ee
According to Eq.~(\ref{cctensor}), the conductivity, due to the
dimensional argument, can be approximated as $\sigma \sim g^2T^2/\gamma$
which combined with the estimate (\ref{sigma-est}) provides
\be
\label{gamma-est}
\frac{1}{\gamma} \sim {1 \over g^2 {\rm ln}(1/g) \: T} 
\sim t_{\rm soft} \;,
\ee
where $t_{\rm soft}$ is the characteristic time scale of the of 
parton-parton collisions at momentum transfers of order $g^2 T$
\cite{Arnold:1998cy}.
We also observe that
\be
\label{sigma-D-est}
\frac{\sigma}{\D} \sim g^2 T^2 
\sim m_D^2 \;,
\ee
where $m_D$ is the screening mass. Having these estimates, we first 
note that all modes of charge density longer than the screening length 
are neutralized dominantly by the ohmic currents. However, we can trust 
Eqs.~(\ref{qgp-solution},\ref{col-diff-eq}) only for time intervals longer 
than $1/\gamma$ because the derivation of the color conductivity presented 
in Sec.~\ref{D-sigma-constants} is valid for $t \gtrsim 1/\gamma$. Taking 
into account the estimate (\ref{sigma-est}), we find that all modes of 
charge density  vanish at $t \gtrsim 1/\gamma$. Since the characteristic 
time scale of color dissipation cannot be shorter than $t_{\rm soft}$, 
the whitening of the quark-gluon plasma occurs at $t \sim t_{\rm soft}$. 
At shorter times scales the color density is expected to oscillate. 
We note that in the electromagnetic plasma local charges are neutralized 
very fast, but the currents survive in the system for a long time as the 
plasma is a very good conductor. Although, the quark-gluon plasma is a rather
poor color conductor, the color currents can still persist in the system
significantly longer than the color charge density \cite{Bodeker:1998hm},
as they couple to the non-perturbative chromomagnetic fields.

Since the conductivity is responsible for whitening of the quark-gluon 
plasma in the long-wave limit, we discuss in more detail the equation
\be \label{col-cond-eq}
\big[ D^0 + \sigma \big] \rho (x) = 0 \;,
\ee
which describes how the ohmic currents neutralize the system.
Eq.~(\ref{col-cond-eq}) is solved by
\be \label{col-solution}
 \rho (x) = \Omega (x,x_0) \:  e^{-\sigma t}  \rho_0({\bf x})
\: \Omega (x_0,x) \;,
\ee
where $x \equiv (t, {\bf x})$, $x_0 \equiv (0, {\bf x})$ and
$\Omega (x,x_0)$ is the parallel transporter
\be
\Omega (x,x_0) = {\cal T} {\rm exp}
\Big[ ig \int_0^t dt' A^0(t',{\bf x}) \Big]\;,
\ee
with ${\cal T}$ denoting the time ordering. Observing that
\be
\Big({\partial \over \partial t} - ig A^0(x) \Big)
\Omega (x,x_0) = \Omega (x_0,x)
\Big( \buildrel \leftarrow \over{\partial \over \partial t}
+ ig A^0(x) \Big) = 0 \;,
\ee
one shows by direct calculation that the expression (\ref{col-solution})
solves Eq.~(\ref{col-cond-eq}). Since $\sigma$ has non-diagonal entries,
various colors are coupled to each other in the course of temporal
evolution.


\section{Discussion}


As discussed in our previous paper \cite{Manuel:2003zr}, parton-parton
collisions thermalize the momentum degrees of freedom but they do not
neutralize the local color charges. To whiten the quark-gluon plasma
collective phenomena are required. The local charges generate
chromoelectric fields, which, in turn, induce  color currents. At the 
scale longer than $\sqrt{\D/\sigma}$, these ohmic currents effectively 
neutralize the system, more effectively than the diffusive currents 
caused by the charge density gradients.

The question arises what is the characteristic time scale of momentum
thermalization and that of the plasma whitening. Our analysis implicitly
assumes that the equilibration of momentum is much faster than the
neutralization, as we linearize the transport equations around the local
equilibrium distribution functions which cancel the collision terms. In
other words, it is implicitly assumed that the plasma momentum distribution
first reaches its local equilibrium form, and then the system is neutralized.
Unfortunately, we are unable to compute the two scales of interest
as it requires an analysis of parton-parton scattering in a colorful
non-equilibrium configuration. Our choice of local equilibrium configuration 
found in \cite{Manuel:2003zr} is to some extent dictated by technical 
reasons. The local equilibrium distribution functions represent a nontrivial 
colorful configuration that is convenient to compute transport coefficients 
as the collision terms then vanish.

The problem of plasma equilibration is also complicated by the fact
that color collective phenomena are not only responsible for
the whitening but they also contribute to the momentum equilibration.
One of us has argued for long time
\cite{Mrowczynski:qm,Mrowczynski:xv,Mrowczynski:1996vh,Randrup:2003cw},
see also \cite{Romatschke:2003ms,Arnold:2003rq}, that color plasma
instabilities, which occur in anisotropic systems, speed up the momentum
thermalization. If the plasma momentum distribution is strongly elongated
in one direction, as it occurs in heavy-ion collisions, the instabilities
generate momentum in the transverse direction, making the system more
isotropic. However, the instabilities also generate local color charges
that have to be neutralized. Thus, the whole process of equilibration
of the quark-gluon plasma is very complex, and it depends on the plasma
initial state.

The only reliable estimates of the time scales of interest have
been found for the perturbative quark-gluon plasma which is close 
to global equilibrium. To equilibrate the system's momentum degrees 
of freedom, the parton-parton interactions with momentum transfers of 
order $T$ are needed. Such a transfer can be achieved in a single
parton-parton collision or as a cumulative effect of many soft
scatterings. The time scale of such processes is \cite{Arnold:1998cy}
\be
t_{\rm hard} \sim {1 \over g^4 {\rm ln}(1/g) \: T } \;.
\ee
As argued in the previous section, the whitening of the quark-gluon 
plasma occurs at $t \gtrsim t_{\rm soft}$. Thus,  the plasma becomes 
white first and then the momentum degrees of freedom thermalize as
$t_{\rm soft} \ll t_{\rm hard}$. Analogous analysis for a colorful 
background should include the effect of the colored chemical potentials 
that might alter the above picture.

At the end, let us recapitulate our considerations. Within the
QCD transport theory we have found the conductivity and diffusion
coefficients in the colorful equilibrium configuration. While the
diffusion constant is proportional to the unit matrix in color
space, the conductivity coefficient has a nontrivial tensorial
structure. The macroscopic equation describing the temporal
evolution of the color charge density has been derived. Its solution
shows that the ohmic currents dominate whitening of the quark-gluon 
plasma at sufficiently long scale.

\acknowledgements

We thank Peter Arnold for pointing out a wrong time scale estimate
of plasma whitening in the first version of our manuscript. This work 
has been supported in part by MCYT (Spain) under grant FPA2001-3031.

\appendix

\section{}


In this Appendix we argue that the commutators $[D^0,\sigma]$ and
$[{\bf D},\sigma]$ are small in a perturbative regime, but we first show that
$[D^\mu,\sigma] = (\widetilde D^\mu \sigma)$ where $\widetilde D^\mu$
is the covariant derivative of rank 2.

We are interested in the following expression
\be
\label{A1}
{\bf D}_{a a'} (\sigma^{a'b} {\bf E}_b ) =
\left(\delta_{a a'} {\bf \nabla}
+ g f^{aea'} {\bf A}^e \right) (\sigma^{a'b} {\bf E}_b ) \;.
\ee
Because of the antisymmetry of the structure constants $f^{abc}$,
 Eq.~(\ref{A1})
can be rewritten as
\be
\label{A2}
{\bf D}_{a a'} (\sigma^{a'b} {\bf E}_b ) =
{\bf D}_{a a'} (\sigma^{a'b} {\bf E}_b )
+ g f^{b e b'}{\bf A}^e \sigma^{a b'} {\bf E}_b
+ g f^{b e b'}{\bf A}^e \sigma^{a b} {\bf E}_{b'}
= \left(\widetilde {\bf D}^{a a'}_{b b'} \sigma^{a'b'} \right) {\bf E}_b
+ \sigma^{ab} {\bf D}_{b b'} {\bf E}_{b'} \ ,
\ee
where
\be
(\widetilde {\bf D})^{ac}_{bd} \equiv \nabla \delta^{ac} \delta^{bd}
+ g f^{aec} \delta^{bd} \: {\bf A}^e
+ g f^{bed} \delta^{ac} \: {\bf A}^e \;.
\ee
In matrix notation Eq.~(\ref{A2}) gets the form
\be
{\bf D}(\sigma {\bf E}) = \sigma \, {\bf D} {\bf E}
+  (\widetilde {\bf D} \sigma)\, {\bf E} \;,
\ee
and thus
\be
[D^\mu,\sigma] = (\widetilde D^\mu \sigma ) \;.
\ee

We are now going to show that $[D^\mu,\sigma]$ is negligible
when $1/g \gg 1$. We actually demonstrate that $(D^\mu \sigma_q)$,
computed in the fundamental representation, is suppressed
by powers of $g$ when compared to $\sigma_q D^\mu$. The same
analysis can be done in the adjoint representation, and for
the gluons, but it requires tedious manipulations with color
indices.

We first compute  ${\bf D} \sigma_q$.  Because of the local isotropy
of the equilibrium state, we have
\be
\label{A4}
{\bf D} \sigma_q  =
{g^2 \over 3\gamma}
\int {d^3 p \over (2\pi)^3} \, {1 \over E_p}
\, {\bf D} \big( Q^{\rm eq} + \bar Q^{\rm eq} \big)
= {g^2 \over \gamma} \int {d^3 p \over (2\pi)^3} \, { {\bf v}\over E_p}
 ( {\bf v} \cdot {\bf D})
\big(Q^{\rm eq} + \bar Q^{\rm eq} \big) \;,
\ee
where we have used the fact that ${\bf v}^2 = 1$. Using the transport
equation (\ref{q-transport-eq2}), Eq.~(\ref{A4}) is rewritten as
\be
\label{A5}
{\bf D} \sigma_q  =
{g^2 \over \gamma} \int {d^3 p \over (2\pi)^3} {{\bf v}\over E_p} \,
\Big( - D^0 ((Q^{\rm eq} + \bar Q^{\rm eq})
+ {g \over 2}  \{{\bf E}, \nabla_p (Q^{\rm eq} - \bar Q^{\rm eq})\}
+ (L[\delta Q] + \bar L[\delta \bar Q]) \Big) \;.
\ee
The first term in the r.h.s. of Eq.~(\ref{A4}) vanishes because of
local isotropy. The remaining two terms are nonzero but
${\bf D} \sigma_q {\bf E}$ is seen to be smaller than
$\sigma_q {\bf D} {\bf E} $ by at least two powers of $g$.
The terms with the collision terms are even more suppressed.

Let us now discuss $D^0 \sigma_q$. Using the transport equation
(\ref{q-transport-eq2}), one gets
\be
\label{A3}
D^0 \sigma_q = {g^2 \over 3\gamma}
\int {d^3 p \over (2\pi)^3} \, {1 \over E_p}
\Big( - {\bf v} \nabla ( Q^{\rm eq} + \bar Q^{\rm eq} )
+ {g \over 2} \{ {\bf E}, \nabla_p ( Q^{\rm eq} - \bar Q^{\rm eq})\}
+ (L[\delta Q] + \bar L[\delta \bar Q])\Big)\;.
\ee
The first and the second term in the r.h.s. of Eq.~(\ref{A3}) both
vanish because of local isotropy of the equilibrium momentum distribution.
The third term does not vanish but $(D^0 \sigma_q) \rho$ is suppressed
with respect to $\sigma_q D^0 \rho$ by powers of $g$ hidden in $L$ and
$\bar L$.

%
%


\begin{thebibliography}{99}

\bibitem{Ackermann:2000tr}
K.~H.~Ackermann {\it et al.}  [STAR Collaboration],
Phys.\ Rev.\ Lett.\  {\bf 86}, 402 (2001)
[arXiv:nucl-ex/0009011].

\bibitem{Heinz:2001xi}
U.~W.~Heinz and P.~F.~Kolb,
Nucl.\ Phys.\ A {\bf 702}, 269 (2002)
[arXiv:hep-ph/0111075].

\bibitem{Manuel:2003zr}
C.~Manuel and St.~Mr\' owczy\' nski,
Phys.\ Rev.\ D {\bf 68}, 094010 (2003)
[arXiv:hep-ph/0306209].

\bibitem{Heinz:qe}
U.~W.~Heinz,
Annals Phys.\  {\bf 168}, 148 (1986).

\bibitem{Czyz:1986mr}
W.~Czyz and W.~Florkowski,
Acta Phys.\ Polon.\ B {\bf 17}, 819 (1986).

\bibitem{Dyrek:1986vv}
A.~Dyrek and W.~Florkowski,
Phys.\ Rev.\ D {\bf 36}, 2172 (1987).

\bibitem{Mrowczynski:xu}
St.~Mr\'owczy\'nski,
Acta Phys.\ Polon.\ B {\bf 19} (1988) 91.

\bibitem{Mrowczynski:1989bv}
St.~Mr\'owczy\'nski,
Adv.\ Ser.\ Direct.\ High Energy Phys.\  {\bf 6}, 185 (1990).

\bibitem{Selikhov:1993ns}
A.~Selikhov and M.~Gyulassy,
Phys.\ Lett.\ B {\bf 316}, 373 (1993)
[arXiv:nucl-th/9307007].

\bibitem{Selikhov:xn}
A.~V.~Selikhov and M.~Gyulassy,
Phys.\ Rev.\ C {\bf 49} (1994) 1726.

\bibitem{Heiselberg:px}
H.~Heiselberg,
Phys.\ Rev.\ Lett.\  {\bf 72}, 3013 (1994)
[arXiv:hep-ph/9401317].

\bibitem{Bodeker:1998hm}
D.~B\"odeker,
Phys.\ Lett.\ B {\bf 426}, 351 (1998)
[arXiv:hep-ph/9801430].

\bibitem{Bodeker:1999ey}
D.~B\"odeker,
Nucl.\ Phys.\ B {\bf 559}, 502 (1999)
[arXiv:hep-ph/9905239].

\bibitem{Litim:1999ns}
D.~F.~Litim and C.~Manuel,
Phys.\ Rev.\ Lett.\  {\bf 82}, 4981 (1999)
[arXiv:hep-ph/9902430].


\bibitem{Litim:1999id}
D.~F.~Litim and C.~Manuel,
Nucl.\ Phys.\ B {\bf 562}, 237 (1999)
[arXiv:hep-ph/9906210].

\bibitem{Arnold:1998cy}
P.~Arnold, D.~T.~Son and L.~G.~Yaffe,
Phys.\ Rev.\ D {\bf 59}, 105020 (1999)
[arXiv:hep-ph/9810216].

\bibitem{Blaizot:1999xk}
J.~P.~Blaizot and E.~Iancu,
Nucl.\ Phys.\ B {\bf 557}, 183 (1999)
[arXiv:hep-ph/9903389].

\bibitem{Arnold:1999uy}
P.~Arnold and L.~G.~Yaffe,
Phys.\ Rev.\ D {\bf 62}, 125014 (2000)
[arXiv:hep-ph/9912306].

\bibitem{MartinezResco:2000pz}
J.~M.~Martinez Resco and M.~A.~Valle Basagoiti,
Phys.\ Rev.\ D {\bf 63}, 056008 (2001)
[arXiv:hep-ph/0009331].

\bibitem{BGK}
P.L.~Bhatnagar, E.P.~Gross, and M.Krook,
Phys. Rev. {\bf 94}, 511 (1954).

\bibitem{Mrowczynski:qm}
St. Mr\' owczy\' nski,
Phys.\ Lett.\ B {\bf 314}, 118 (1993).

\bibitem{Mrowczynski:xv}
St. Mr\' owczy\' nski,
Phys.\ Rev.\ C {\bf 49}, 2191 (1994).

\bibitem{Mrowczynski:1996vh}
St. Mr\' owczy\' nski,
Phys.\ Lett.\ B {\bf 393}, 26 (1997).

\bibitem{Randrup:2003cw}
J.~Randrup and St.~Mr\'owczy\'nski,
Phys.\ Rev.\ C {\bf 68}, 034909 (2003)
[arXiv:nucl-th/0303021].

\bibitem{Romatschke:2003ms}
P.~Romatschke and M.~Strickland,
Phys.\ Rev.\ D {\bf 68}, 036004 (2003)
[arXiv:hep-ph/0304092].

\bibitem{Arnold:2003rq}
P.~Arnold, J.~Lenaghan and G.~D.~Moore,
JHEP {\bf 0308}, 002 (2003)
[arXiv:hep-ph/0307325].

\end{thebibliography}
\end{document}